\documentclass{nature}

\usepackage{color}
\usepackage{times}
\usepackage{graphicx}
\usepackage{comment}
\usepackage[intlimits]{amsmath}

\newcommand{\ndfrac}[2]{\frac{\text{d} #1}{\text{d} #2}}

\title{\textcolor{black}{Targeted Assembly and Synchronization of Self-Spinning Microgears}}

\author
{Antoine Aubret$^{1}$, Mena Youssef $^{2}$,  Stefano Sacanna$^{2}$, J\'er\'emie Palacci $^{1\ast}$ \\
\normalsize{$^{1}$Department of Physics, University of California, San Diego, USA}\\
\normalsize{$^{2}$Department of Chemistry, New York University,  USA}\\
}

\date{}

\begin{document} 


\baselineskip24pt


\maketitle


\begin{abstract}
Self-assembly is the autonomous organization of components into patterns or structures:  an essential ingredient of biology and a desired route to  complex organization\cite{Whitesides:2002uq}. At equilibrium, the structure is encoded through specific interactions\cite{Erb:2009cz, Sacanna:2010bv, Chen:2011be, Wang:2012gd, Damasceno:2012gi, Manoharan:2015ko, Ducrot:2017cs},  at an unfavorable entropic cost for the system.  An alternative approach, widely used by Nature,  uses energy input to bypass the entropy bottleneck and develop features otherwise impossible at equilibrium\cite{Needleman:2017fq}. Dissipative building blocks that inject energy locally were made available by recent advance in colloidal science \cite{Dey:2016ika, Aubret:2017ch} but have \textcolor{black}{not been used to control self-assembly. Here we show the robust formation of self-powered rotors and dynamical superstructures from active particles and harness non-equilibrium phoretic phenomena\cite{anderson} to tailor interactions and direct self-assembly.  We use a photoactive component that consumes fuel, hematite,  to devise phototactic microswimmers that form self-spinning microgears following spatiotemporal light patterns. The gears  are coupled {\it via} their chemical clouds  and constitute the elementary bricks of synchronized superstructures, which autonomously regulate their dynamics. The results are quantitatively rationalized on the basis of a stochastic description of  diffusio-phoretic oscillators dynamically coupled by chemical gradients to form directional interactions. 
Our findings demonstrate that non-equilibrium phenomena can be harnessed to shape interactions and program hierarchical constructions. It lays the groundwork for the self-assembly of dynamical architectures and synchronized micro-machinery.}
\end{abstract}

The self-assembly of complex structures  which emulate the fidelity and tunability of their biological counterparts is a fundamental goal of material science and engineering. In colloidal science, the paradigm for equilibrium self-assembly encodes information through specific interactions between building blocks that, in turn, promote assembly, so that the resulting (static) structure is the thermodynamic ground state. Emergent properties in Active Matter\cite{Ramaswamy:2010bf}, in which self-driven units convert an energy source into useful motion and work,  have been the focus of intensive experimental and theoretical studies in recent years and primarily focused on meso-scale or macroscopic fluxes\cite{Cates:2010um}, flocks\cite{Bricard:2014jq,Vicsek:1995fk} or flows\cite{Wu:2017ct, Lushi:2014fn} from local mechanical forcing. Their use to design nonequilibrium interactions  and control self-assembly  remains, however, largely unexplored\cite{Wang:2013dv, Banerjee:2016ce,ShieldsIV:2017jf}. We introduce a sequential approach in order to architect structures, the dynamics of which arise from chemical gradients through diffusiophoretic interactions.  The process is  dissipative and utilizes a photocatalytic material, hematite, to harvest energy from a hydrogen peroxide fuel\textcolor{black}{\cite{Palacci:2013eu} and form hierarchical  superstructures from elementary bricks, through non-equilibrium pathways (Fig.1).} \textcolor{black}{Instrumental to our work is the development of phototactic phototactic swimmers (Fig.1A)}, which  direct along light gradients and assemble into self-spinning microgears or {\it rotors} (Fig.1B). 
We focus in this Letter on the self-organization of moderate numbers of rotors with sorted spins. They organize into stable patterns, which dynamics originates from the interplay between the phase and the space coordinates of the components. For example, sets of 7 rotors form  hexagonal patterns (movie S1, Fig.1C), and exhibit an edge-current traveling at  $\Omega \sim 0.1$ rad/s for co-rotating rotors  (movie S2) or remain static for structures with alternating spins  (movie S3). Switching off the light, the system swiftly returns to equilibrium, the interactions vanish and the order is destroyed by thermal noise (Fig.1C-inset). 

A hematite cube ($\alpha-$Fe$_2$O$_3$)  decomposes hydrogen peroxide, at rate $\nu=\beta I/({I+I_0})$, dependent on the light intensity $I$\cite{Palacci:2013eu}.  It travels toward the low intensity in a gradient as a result of the asymmetric fuel consumption and diffusiophoresis\cite{anderson},  the migration of a colloidal particle in a concentration gradient $\nabla c$.  The measured velocity shows a good agreement with the diffusiophoretic velocity, $V\propto \nabla c\propto \nabla \nu(I)$, predicted from  the experimental intensity profile (Fig.2A). The migration is slow, $V\sim2$ $\mu$m/s, as the gradient on the particle scale is moderate. We take advantage of this feature to \textcolor{black}{devise} phototactic swimmers, with a fore-aft asymmetry and consisting of the extrusion of the hematite cube  from a chemically inert polymer bead\cite{Youssef:2016kb} (SI). They exhibit a persistent random walk in uniform light, the bead heading, with a velocity  up to $30$ $\mu$m/s, controlled by the intensity. In a light gradient, they direct towards the high intensity following the reorientation by the hematite component. We use a custom optical setup combining light sources  to deliver spatio-temporal patterns  with   a few microns and  $0.01$s accuracy   and guide the self-assembly (SI). 

A dilute suspension of the swimmers is fed to a capillary, residing near the bottom surface at surface density $\Phi_s\sim 10^{-3}$ part$/\mu$m$^2$. They travel  isotropically in uniform light but gather and collide as we superimpose the bright spot of a focused laser ($\lambda=404$ nm)\textcolor{black}{, assembling into a self-spinning microgear. The structure is composed of a \textcolor{black}{core} vertical swimmer surrounded by  6 peripheral close-packed particles (Fig.1B). The rotor remains stable  in uniform light  after extinction of the laser spot and its lifetime is  set by the duration of the experiment ($\sim10-20$ mins).  We repeat the light sequence to form multiple rotors with high yield and control. \textcolor{black}{The rotor constitutes a "ground state" as added particles spontaneously detach} (movie S4). The handedness is random and we measure $49.85\%$  clockwise and  $50.15\%$ counter-clockwise {\bf from 1017 rotors}.\\ As hematite is partially absorbing at the wavelength of activation, it sets a gradient of light and reaction rate along the illumination axis. The effect is negligible in most situations but becomes significant as a swimmer crosses the brightest region of a laser spot: for a laser shined from below, the hematite is phoretically lifted,  and the swimmer flips vertically. It swims downwards against the bottom wall,  producing a hydrodynamic pumping  \cite{Weinert:2008cn, DiLeonardo:ul} that attracts neighboring swimmers and drives the assembly into the self-spinning structure \textcolor{black}{(Fig.1D)}.}
 In our experimental conditions, we do not observe decomposition of the hydrogen peroxide nor optical forces from the light. Shining the laser from the top does not allow the vertical flipping of a swimmer and inhibits the formation of the rotors. Transient rotors can be formed focusing the laser on a sphere deposited on the substrate, however with limited  yield and lifetime, which stresses the importance of the hydrodynamic pumping for the cohesion of the structure. The angular speed $\omega$ is tuned by the light intensity (Fig.2B), \textcolor{black}{reflecting the translational velocity of the individual swimmers (Fig. 2B) and shows  less than $10\%$ variability amongst a population}.  At low  speeds, $\omega<3$ rad/s,  fluctuations can flip the direction of rotation, though the magnitude of the rotation rate is unchanged (Fig.2B-inset). At higher speeds, unidirectional motion persists over the time of the experiment: the non-equilibrium forces stabilize the structure (Fig.2B-inset). In uniform light, the center of mass of the rotors displays a two-dimensional random walk with a diffusion coefficient  $D_\mathrm{R}= 0.4\pm 0.1$ $\mu$m$^2$/s, larger than a passive particle of comparable size. The rotors show slow migration in light gradients, which we harness to build superstructures. 

We probe the influence of a rotor \textcolor{black}{on its surrounding} using fluorescent tracers (latex, 200 nm, diffusivity $D_\text{c}$), which concentration $\rho(r)$  is extracted by fluorescence microscopy and azimuthal average (Fig.2C-inset). We observe a radial repulsion  from the rotor, constituting a sink of hydrogen peroxide, with $c\propto 1/r$.  The tracers migrate in the concentration gradient by diffusiophoresis, with velocity $V_{DP}\propto \nabla c = \alpha/r^2,$ ($\alpha>0$ for a repulsive interaction). \textcolor{black}{ At steady state,  the flux of particles $j = \rho V_{\text{DP}} - D_\text{c} \nabla_r \rho$ gives a Boltzmann distribution, $\rho \propto e^{-\frac{\alpha}{D_\text{c}r}}$, in an effective repulsive potential,   in agreement with the experiment for  $\alpha=48\pm10$ $\mu$m$^3$/s (Fig.2C).}\\
We now discuss the interactions between pairs of rotors. Rotors radially repel each other as previously observed  for latex particles. The radial repulsion is isotropic and insensitive to the absolute or relative handedness of the rotors. Using isotropic light patterns with a flat central region, we control the confinement of the rotors and observe the existence of short-range tangential interaction for $r\sim2.0R-3.0R$, where $r$ is the center-to-center distance of rotors of radii $R\sim 3$ $\mu$m.  Co-rotating pairs revolve around each other at $\Omega\sim 0.05 $ rad/s, in the direction of their spin (movie S5, Fig.3A) but do not synchronize (Fig.3B). Counter-rotating pairs are static (Fig.3A, movie S6) and phase-lock at $\pi/6$  as visible from the peak in the Probability Density Function (PDF) of the phase lag $\Psi =|\theta_1|-|\theta_2|$ (Fig.3B). The PDF broadens as the rotors are further apart and becomes asymmetric for pairs with different angular speeds (Fig.3B-inset). This behavior is reminiscent of mechanical cogwheels, in the absence of any contact between the rotors.  We rule out hydrodynamics as  the main contribution to the tangential interaction, as a torque-free rotor generates a slip-flow opposite to its spin, which makes a co-rotating pair revolve in that direction, in contrast with the experiment.

In order to gain insight into the diffusiophoretic coupling, we study the concentration field surrounding a rotor and simulate the diffusion equation for a structure of 7 impermeable and passive spheres, decorated by 6 chemically active sites (see SI for details). It constitutes a sink of fuel, which near-field concentration field presents the 6-fold symmetry of the rotor (Fig.3C). 
We compute, from the simulated concentration field, the radial, respectively azimuthal, phoretic velocities for a point particle $\tilde v_{r,\theta} \propto \nabla_{r,\theta}  c(r,\theta)$  and obtain $1/r^2$ decay, respectively  $1/r^8$,  with $(\tilde v_{\theta}/\tilde v_{r})\cdot {(r/R)}^6\sim1$ (Fig.3D). 
 \textcolor{black}{The short-range of the tangential interaction originates from the rapid decay of the azimuthal component. It arises from the superimposition of the monopolar concentration field generated by each active site and reflects the azimuthal dependence of the concentration surrounding a sphere with hexapolar chemical activity: $c(r,\theta)\propto \mathrm{P}_6(\cos\theta)/r^7$, where $P_6$ is the Legendre polynomial of $6^{th}$ degree \cite{Golestanian:2007hu}}. The result is qualitatively unchanged for a bead of finite size (Fig. 3D), which velocity is obtained by integration of the slip on the surface  \cite{anderson}, or by addition of an impermeable wall delineating a semi-infinite space (see SI).  

As a result of the fast diffusion of  the fuel, $D_\text{H$_2$O$_2$}\sim 10^{3}$ $\mu$m$^2$/s, the P\'eclet number of the rotation is low, $\text{Pe}=R^2\omega/D_\text{H$_2$O$_2$}\sim 0.01$,  and the concentration field is steady in the rotating frame of the rotor. Neighboring rotors interact through the modulation of the concentration field computed above, mirroring the effective  energy landscape $U(r,\theta) \propto c(r,\theta)$\cite{Banerjee:2016ce}.  The tangential interaction is short-range and we assume that the azimuthal coupling is set at distance $r-R$, separating the edge of a rotor to the center to its neighbor. The phases $\theta_i$ (i=1,2)  of the rotors in a pair, at fixed distance $r$, are described by coupled Langevin equations $\text{d}\theta_i/\text{d}t = \omega_i + \varepsilon Q(\theta_i, \theta_j) + \sqrt{2D_\theta} \zeta(t)$, 
where $i,j$ identifies the rotor, $\omega_i$ their angular speed,  $\varepsilon Q$ is the phoretic coupling between the rotors  and $\zeta(t)$ a delta-correlated noise of  amplitude $D_\theta$. We \textcolor{black}{independently estimate the amplitude of the phoretic coupling  from the radial  repulsion}, $\varepsilon=\frac{\alpha}{R^3} \cdot {\left(\frac{R}{r-R}\right)}^8 \sim 2 {\left(\frac{R}{r-R}\right)}^8$. Guided by the numerical results, we  approximate the coupling term, $Q=\sin(6 \Psi)$, with the phase-lag $\Psi=|\theta_1|-|\theta_2|$. We \textcolor{black}{finally} obtain by summation: 
 $$\ndfrac{\Psi}{t} = -\delta \omega +2\varepsilon\sin(6\Psi) + \sqrt{4D_\theta}\zeta(t)$$
where  $\delta \omega$ is the relative  speed of the rotors (see SI for details). This is a classical Adler equation of synchronization with noise, also seen as the dynamics of an overdamped  Brownian particle in a tilted Washboard potential, $\mathcal{V}(\Psi)=\delta \omega \Psi + (\varepsilon/3) \cos (6\Psi)$\cite{R:1973wu,DiLeonardo:2012}. In the absence of noise, the phase is trapped in a potential well  at high coupling, $3\delta \omega/\varepsilon <1$, and slides down the corrugated energy landscape at low coupling  \cite{Shlomovitz:2014gv}. This qualitatively describes the  phase-locking observed for contra-rotating rotors, for which $\delta \omega\sim 0$, and the slow  revolution at $\Omega$ of co-rotating pairs, for which $\delta \omega\sim 2\omega$. In the presence of noise, the PDF for $\Psi$ follows a Fokker-Planck equation, which  stationary solution is:
$P(\Psi)=(1/A) \text{e}^{-\mathcal{V}(\Psi)/2D_\theta} \int_\Psi^{\Psi+\pi/3}{\text{e}^{\mathcal{V}(\Psi')/2D_\theta} d\Psi'}$, where A is the normalization constant. It  shows an excellent  agreement with our experimental measurements of the PDF for the phase lag (Fig.3B), with \textcolor{black}{the phoretic coupling}  $\varepsilon$ as single fit parameter, as we independently extract the rotation speeds and noise amplitude, $D_\theta=0.03 \pm 0.01$ rad$^2$/s  from individual tracking of the rotation of the rotors (see SI).  The model remarkably captures the asymmetric PDF observed for pairs of contra-rotating rotors with notable speed difference without additional parameters (Fig.3B-inset). 
The coupling parameter $\varepsilon$ decays with increasing distance and shows a good agreement with \textcolor{black}{our prediction, }$\varepsilon \sim 2 {\left(\frac{R}{r-R}\right)}^8$, \textcolor{black}{obtained from the radial repulsion between rotors, without adjustable parameters  (Fig.3E)}.  For co-rotating pairs, guided by the persistence of the phase lag, we neglect the noise and the difference of rotation speed of the rotors, and obtain $\text{d}\theta/\text{d}t = \omega + \varepsilon \sin(12\theta)$, predicting a reduced rotation rate  $ \omega\sqrt{1-(\varepsilon/\omega)^2}$ (see SI). The short-range of the tangential interaction confines the slowing down to the nearest parts of rotors, effectively acting as partial friction. It qualitatively agrees with the experiment and rotors revolve in the direction of their spin, with $\tilde\Omega\sim\omega -  \omega\sqrt{1-(\varepsilon/\omega)^2}\sim \frac{\varepsilon^2}{2\omega}\sim 0.005$ rad/s.  \\
Following our understanding of the pair interaction, we architect machineries, whose collective dynamics arise from the spin sequence of their components.  First, we control the travel speed $\Omega$ of the edge-current  in hexagonal patterns of 7 co-rotating gears (Fig.1C, 4A), increasing the azimuthal coupling $\varepsilon$ through  confinement (Fig. 4A-inset). Next, we form higher-order assemblies  by combination of superstructures: we initiate two contra-rotating sets of 3 co-rotating rotors, each collectively rotating along their common direction of spin (Fig.4B). They are subsequently combined to constitute the synchronized gears of a micromachine (movie S7, Fig.4C), stressing the robustness and versatility of \textcolor{black}{our findings}.  

\textcolor{black}{In summary, we demonstrate a novel type of self-assembly that uses dissipative building blocks to achieve functioning micromachines. We engineer active particles that auto-organize into self-spinning microgears with  anisotropic interactions. They result into the dynamical synchronization of  pairs of rotors, that we characterize and quantitatively rationalize with a stochastic description of phoretic oscillators coupled by chemical gradients. Assembly of those rotors exhibit collective dynamics, encoded in the spatial sequence of their components. The interplay between phase synchronization and spatial organization has the potential to achieve new form of self-organization, without equilibrium counterparts nor observed for collections of {\it translational} self-propelled particles\cite{vanZuiden:2016ek}. Self-powered microgears are moreover the key components of geared topological metamaterials, the mechanical counterpart of electronic topological insulators\cite{Meeussen:2016dl}. Our bottom-up approach combines  the tailoring of directional non-equilibrium interactions with spatiotemporal sequences of light to program dynamical superstructures that autonomously regulate. It makes us architect of Matter and brings biological organization to the material world.}

\section*{References}
\bibliographystyle{naturemag}

\begin{thebibliography}{10}
\expandafter\ifx\csname url\endcsname\relax
  \def\url#1{\texttt{#1}}\fi
\expandafter\ifx\csname urlprefix\endcsname\relax\def\urlprefix{URL }\fi
\providecommand{\bibinfo}[2]{#2}
\providecommand{\eprint}[2][]{\url{#2}}

\bibitem{Whitesides:2002uq}
\bibinfo{author}{Whitesides, G.~M.} \& \bibinfo{author}{Grzybowski, B.}
\newblock \bibinfo{title}{{Self-assembly at all scales}}.
\newblock \emph{\bibinfo{journal}{Science}} \textbf{\bibinfo{volume}{295}},
  \bibinfo{pages}{2418--2421} (\bibinfo{year}{2002}).

\bibitem{Erb:2009cz}
\bibinfo{author}{Erb, R.~M.}, \bibinfo{author}{Son, H.~S.},
  \bibinfo{author}{Samanta, B.}, \bibinfo{author}{Rotello, V.~M.} \&
  \bibinfo{author}{Yellen, B.~B.}
\newblock \bibinfo{title}{{Magnetic assembly of colloidal superstructures with
  multipole symmetry}}.
\newblock \emph{\bibinfo{journal}{Nature}} \textbf{\bibinfo{volume}{457}},
  \bibinfo{pages}{999--1002} (\bibinfo{year}{2009}).

\bibitem{Sacanna:2010bv}
\bibinfo{author}{Sacanna, S.}, \bibinfo{author}{Irvine, W. T.~M.},
  \bibinfo{author}{Chaikin, P.~M.} \& \bibinfo{author}{Pine, D.~J.}
\newblock \bibinfo{title}{{Lock and key colloids}}.
\newblock \emph{\bibinfo{journal}{Nature}} \textbf{\bibinfo{volume}{464}},
  \bibinfo{pages}{575--578} (\bibinfo{year}{2010}).

\bibitem{Chen:2011be}
\bibinfo{author}{Chen, Q.}, \bibinfo{author}{Bae, S.~C.} \&
  \bibinfo{author}{Granick, S.}
\newblock \bibinfo{title}{{Directed self-assembly of a colloidal kagome
  lattice}}.
\newblock \emph{\bibinfo{journal}{Nature}} \textbf{\bibinfo{volume}{469}},
  \bibinfo{pages}{381--384} (\bibinfo{year}{2011}).

\bibitem{Wang:2012gd}
\bibinfo{author}{Wang, Y.}, \bibinfo{author}{Wang, Y.}, \bibinfo{author}{Breed, D.},  \bibinfo{author}{Manoharan, V.N.} \bibinfo{author}{Feng, L.}, \bibinfo{author}{Hollingsworth, A.D.}, \bibinfo{author}{Weck, M.}, \bibinfo{author}{Pine, D.J.},
\newblock \bibinfo{title}{{Colloids with valence and specific directional
  bonding}}.
\newblock \emph{\bibinfo{journal}{Nature}} \textbf{\bibinfo{volume}{490}},
  \bibinfo{pages}{51--55} (\bibinfo{year}{2012}).

\bibitem{Damasceno:2012gi}
\bibinfo{author}{Damasceno, P.~F.}, \bibinfo{author}{Engel, M.} \&
  \bibinfo{author}{Glotzer, S.~C.}
\newblock \bibinfo{title}{{Predictive Self-Assembly of Polyhedra into Complex
  Structures}}.
\newblock \emph{\bibinfo{journal}{Science}} \textbf{\bibinfo{volume}{337}},
  \bibinfo{pages}{453--457} (\bibinfo{year}{2012}).

\bibitem{Manoharan:2015ko}
\bibinfo{author}{Manoharan, V.~N.}
\newblock \bibinfo{title}{{Colloidal matter: Packing, geometry, and entropy}}.
\newblock \emph{\bibinfo{journal}{Science}} \textbf{\bibinfo{volume}{349}},
  \bibinfo{pages}{942} (\bibinfo{year}{2015}).

\bibitem{Ducrot:2017cs}
\bibinfo{author}{Ducrot, {\'E}.}, \bibinfo{author}{He, M.},
  \bibinfo{author}{Yi, G.-R.} \& \bibinfo{author}{Pine, D.~J.}
\newblock \bibinfo{title}{{Colloidal alloys with preassembled clusters
  and~spheres}}.
\newblock \emph{\bibinfo{journal}{Nature Materials}}
  \textbf{\bibinfo{volume}{16}}, \bibinfo{pages}{652--657}
  (\bibinfo{year}{2017}).

\bibitem{Needleman:2017fq}
\bibinfo{author}{Needleman, D.} \& \bibinfo{author}{Dogic, Z.}
\newblock \bibinfo{title}{{Active matter at the interface between materials
  science and cell biology}}.
\newblock \emph{\bibinfo{journal}{Nature Reviews Materials}}
  \textbf{\bibinfo{volume}{2}}, \bibinfo{pages}{17048--14}
  (\bibinfo{year}{2017}).

\bibitem{Dey:2016ika}
\bibinfo{author}{Dey, K.~K.}, \bibinfo{author}{Wong, F.},
  \bibinfo{author}{Altemose, A.} \& \bibinfo{author}{Sen, A.}
\newblock \bibinfo{title}{{Catalytic Motors-Quo Vadimus}}.
\newblock \emph{\bibinfo{journal}{Current Opinion in Colloid {\&} Interface
  Science}} \textbf{\bibinfo{volume}{21}}, \bibinfo{pages}{4--13}
  (\bibinfo{year}{2016}).

\bibitem{Aubret:2017ch}
\bibinfo{author}{Aubret, A.}, \bibinfo{author}{Ramananarivo, S.} \&
  \bibinfo{author}{Palacci, J.}
\newblock \bibinfo{title}{{Eppur si muove, and yet it moves: Patchy (phoretic)
  swimmers}}.
\newblock \emph{\bibinfo{journal}{Current Opinion in Colloid {\&} Interface
  Science}} \textbf{\bibinfo{volume}{30}}, \bibinfo{pages}{81--89}
  (\bibinfo{year}{2017}).

\bibitem{anderson}
\bibinfo{author}{Anderson, J.~L.}
\newblock \bibinfo{title}{{Colloid Transport by Interfacial Forces}}.
\newblock \emph{\bibinfo{journal}{Annual Review Of Fluid Mechanics}}
  \textbf{\bibinfo{volume}{21}}, \bibinfo{pages}{61--99}
  (\bibinfo{year}{1989}).

\bibitem{Ramaswamy:2010bf}
\bibinfo{author}{Ramaswamy, S.}
\newblock \bibinfo{title}{{The Mechanics and Statistics of Active Matter}}.
\newblock \emph{\bibinfo{journal}{Annual Review of Condensed Matter Physics,
  Vol 1}} \textbf{\bibinfo{volume}{1}}, \bibinfo{pages}{323--345}
  (\bibinfo{year}{2010}).

\bibitem{Cates:2010um}
\bibinfo{author}{Cates, M.~E.}, \bibinfo{author}{Marenduzzo, D.},
  \bibinfo{author}{Pagonabarraga, I.} \& \bibinfo{author}{Tailleur, J.}
\newblock \bibinfo{title}{{Arrested phase separation in reproducing bacteria
  creates a generic route to pattern formation}}.
\newblock \emph{\bibinfo{journal}{Proceedings of the National Academy of
  Sciences of the U.S.A}} \textbf{\bibinfo{volume}{107}},
  \bibinfo{pages}{11715--11720} (\bibinfo{year}{2010}).

\bibitem{Bricard:2014jq}
\bibinfo{author}{Bricard, A.}, \bibinfo{author}{Caussin, J.-B.},
  \bibinfo{author}{Desreumaux, N.}, \bibinfo{author}{Dauchot, O.} \&
  \bibinfo{author}{Bartolo, D.}
\newblock \bibinfo{title}{{Emergence of macroscopic directed motion in
  populations of motile colloids}}.
\newblock \emph{\bibinfo{journal}{Nature}} \textbf{\bibinfo{volume}{503}},
  \bibinfo{pages}{95--98} (\bibinfo{year}{2014}).

\bibitem{Vicsek:1995fk}
\bibinfo{author}{Vicsek, T.}, \bibinfo{author}{Czirok, A.},
  \bibinfo{author}{Benjacob, E.}, \bibinfo{author}{Cohen, I.} \&
  \bibinfo{author}{Shochet, O.}
\newblock \bibinfo{title}{{Novel Type of Phase-Transition in a System of
  Self-Driven Particles}}.
\newblock \emph{\bibinfo{journal}{Physical Review Letters}}
  \textbf{\bibinfo{volume}{75}}, \bibinfo{pages}{1226--1229}
  (\bibinfo{year}{1995}).

\bibitem{Wu:2017ct}
\bibinfo{author}{Wu, K.-T.}, \bibinfo{author}{Hishamunda, J. B.}, \bibinfo{author}{Chen, D.T. N.}, \bibinfo{author}{DeCamp, S.J.}, \bibinfo{author}{Chang, Y.-W.}, \bibinfo{author}{Fernandez-Nieves, A.}, \bibinfo{author}{Fraden, S.}, \bibinfo{author}{Dogic, Z.},
\newblock \bibinfo{title}{{Transition from turbulent to coherent flows in
  confined three-dimensional active fluids}}.
\newblock \emph{\bibinfo{journal}{Science}} \textbf{\bibinfo{volume}{355}},
  \bibinfo{pages}{eaal1979} (\bibinfo{year}{2017}).

\bibitem{Lushi:2014fn}
\bibinfo{author}{Lushi, E.}, \bibinfo{author}{Wioland, H.} \&
  \bibinfo{author}{Goldstein, R.~E.}
\newblock \bibinfo{title}{{Fluid flows created by swimming bacteria drive
  self-organization in confined suspensions}}.
\newblock \emph{\bibinfo{journal}{Proceedings of the National Academy of
  Sciences of the U.S.A}} \textbf{\bibinfo{volume}{111}},
  \bibinfo{pages}{9733--9738} (\bibinfo{year}{2014}).

\bibitem{Wang:2013dv}
\bibinfo{author}{Wang, W.}, \bibinfo{author}{Duan, W.}, \bibinfo{author}{Sen,
  A.} \& \bibinfo{author}{Mallouk, T.~E.}
\newblock \bibinfo{title}{{Catalytically powered dynamic assembly of rod-shaped
  nanomotors and passive tracer particles}}.
\newblock \emph{\bibinfo{journal}{Proceedings Of The National Academy Of
  Sciences Of The United States Of America}} \textbf{\bibinfo{volume}{110}},
  \bibinfo{pages}{17744--17749} (\bibinfo{year}{2013}).

\bibitem{Banerjee:2016ce}
\bibinfo{author}{Banerjee, A.}, \bibinfo{author}{Williams, I.},
  \bibinfo{author}{Azevedo, R.~N.}, \bibinfo{author}{Helgeson, M.~E.} \&
  \bibinfo{author}{Squires, T.~M.}
\newblock \bibinfo{title}{{Soluto-inertial phenomena: Designing long-range,
  long-lasting, surface-specific interactions in suspensions}}.
\newblock \emph{\bibinfo{journal}{Proceedings of the National Academy of
  Sciences of the U.S.A}} \textbf{\bibinfo{volume}{113}},
  \bibinfo{pages}{8612--8617} (\bibinfo{year}{2016}).

\bibitem{ShieldsIV:2017jf}
\bibinfo{author}{Shields~IV, C.~W.} \& \bibinfo{author}{Velev, O.~D.}
\newblock \bibinfo{title}{{The Evolution of Active Particles: Toward Externally
  Powered Self-Propelling and Self-Reconfiguring Particle Systems}}.
\newblock \emph{\bibinfo{journal}{CHEMPR}} \textbf{\bibinfo{volume}{3}},
  \bibinfo{pages}{539--559} (\bibinfo{year}{2017}).

\bibitem{Vutukuri:2017gx}
\bibinfo{author}{Vutukuri, H.~R.}, \bibinfo{author}{Bet, B.},
  \bibinfo{author}{Roij, R.~x.}, \bibinfo{author}{Dijkstra, M.} \&
  \bibinfo{author}{Huck, W. T.~S.}
\newblock \bibinfo{title}{{Rational design and dynamics of self-propelled
  colloidal bead chains: from rotators to flagella}}.
\newblock \emph{\bibinfo{journal}{Scientific Reports}} \textbf{\bibinfo{volume}{7}} \bibinfo{pages}{1--14}
  (\bibinfo{year}{2017}).

\bibitem{Palacci:2013eu}
\bibinfo{author}{Palacci, J.}, \bibinfo{author}{Sacanna, S.},
  \bibinfo{author}{Steinberg, A.~P.}, \bibinfo{author}{Pine, D.~J.} \&
  \bibinfo{author}{Chaikin, P.~M.}
\newblock \bibinfo{title}{{Living Crystals of Light-Activated Colloidal
  Surfers}}.
\newblock \emph{\bibinfo{journal}{Science}} \textbf{\bibinfo{volume}{339}},
  \bibinfo{pages}{936--940} (\bibinfo{year}{2013}).

\bibitem{Youssef:2016kb}
\bibinfo{author}{Youssef, M.}, \bibinfo{author}{Hueckel, T.},
  \bibinfo{author}{Yi, G.-R.} \& \bibinfo{author}{Sacanna, S.}
\newblock \bibinfo{title}{{Shape-shifting colloids via stimulated dewetting}}.
\newblock \emph{\bibinfo{journal}{Nature Communications}}
  \textbf{\bibinfo{volume}{7}}, \bibinfo{pages}{1--7} (\bibinfo{year}{2016}).

\bibitem{Howse:2007ed}
\bibinfo{author}{Howse, J.~R.}, \bibinfo{author}{Richard A.L.}, \bibinfo{author}{Ryan, A.J.}, \bibinfo{author}{Gough, T.}, \bibinfo{author}{Vafabakhsh, R.}, \bibinfo{author}{Golestanian, R.}
\newblock \bibinfo{title}{{Self-motile colloidal particles: From directed
  propulsion to random walk}}.
\newblock \emph{\bibinfo{journal}{Physical Review Letters}}
  \textbf{\bibinfo{volume}{99}}, \bibinfo{pages}{048102--5} (\bibinfo{year}{2007}).

\bibitem{Maggi:2015dx}
\bibinfo{author}{Maggi, C.}, \bibinfo{author}{Simmchen, J.}, \bibinfo{author}{Saglimbeni, F.}, \bibinfo{author}{Katuri, J.}, \bibinfo{author}{Dipalo, M.}, \bibinfo{author}{De Angelis, F.}, \bibinfo{author}{Sanchez, S.}, \bibinfo{author}{Di Leonardo, R.},
\newblock \bibinfo{title}{{Self-Assembly of Micromachining Systems Powered by
  Janus Micromotors}}.
\newblock \emph{\bibinfo{journal}{Small}} \textbf{\bibinfo{volume}{12}},
  \bibinfo{pages}{446--451} (\bibinfo{year}{2015}).

\bibitem{Weinert:2008cn}
\bibinfo{author}{Weinert, F.~M.} \& \bibinfo{author}{Braun, D.}
\newblock \bibinfo{title}{{Observation of Slip Flow in Thermophoresis}}.
\newblock \emph{\bibinfo{journal}{Physical Review Letters}}
  \textbf{\bibinfo{volume}{101}}, \bibinfo{pages}{168301--4 }
  (\bibinfo{year}{2008}).

\bibitem{DiLeonardo:ul}
\bibinfo{author}{Di~Leonardo, R.}, \bibinfo{author}{Ianni, F.} \&
  \bibinfo{author}{Ruocco, G.}
\newblock \bibinfo{title}{{Colloidal Attraction Induced by a Temperature
  Gradient}}.
\newblock \emph{\bibinfo{journal}{Langmuir}} \textbf{\bibinfo{volume}{25}},
  \bibinfo{pages}{4247--4250} (\bibinfo{year}{2009}).

\bibitem{Leoni:2011gh}
\bibinfo{author}{Leoni, M.} \& \bibinfo{author}{Liverpool, T.~B.}
\newblock \bibinfo{title}{{Dynamics and interactions of active rotors}}.
\newblock \emph{\bibinfo{journal}{Europhysics Letters}}
  \textbf{\bibinfo{volume}{92}}, \bibinfo{pages}{64004--7}
  (\bibinfo{year}{2011}).

\bibitem{Golestanian:2007hu}
\bibinfo{author}{Golestanian, R.}, \bibinfo{author}{Liverpool, T.~B.} \&
  \bibinfo{author}{Ajdari, A.}
\newblock \bibinfo{title}{{Designing phoretic micro- and nano-swimmers}}.
\newblock \emph{\bibinfo{journal}{New Journal of Physics}}
  \textbf{\bibinfo{volume}{9}}, \bibinfo{pages}{126-135} (\bibinfo{year}{2007}).

\bibitem{R:1973wu}
\bibinfo{author}{Adler, R.}
\newblock \bibinfo{title}{{Locking Phenomena in Oscillators}}.
\newblock \emph{\bibinfo{journal}{Proceedings of the IEEE}}
  \textbf{\bibinfo{volume}{61}}, \bibinfo{pages}{1380--1385}
  (\bibinfo{year}{1973}).

\bibitem{DiLeonardo:2012}
\bibinfo{author}{Di Leonardo, R.}, \bibinfo{author}{B\'uz\'as, A.},
  \bibinfo{author}{Kelemen, L.}, \bibinfo{author}{Vizsnyiczai, G.}, \bibinfo{author}{Oroszi, L.} \&
  \bibinfo{author}{Ormos, P.}
\newblock \bibinfo{title}{{Hydrodynamic Synchronization of Light Driven Microrotors}}.
\newblock \emph{\bibinfo{journal}{Physical Review Letters}}
  \textbf{\bibinfo{volume}{109}}, \bibinfo{pages}{034104}
  (\bibinfo{year}{2012}).

\bibitem{Shlomovitz:2014gv}
\bibinfo{author}{Shlomovitz, R.}, \bibinfo{author}{Roongthumskul, Y.},
  \bibinfo{author}{Ji, S.}, \bibinfo{author}{Bozovic, D.} \&
  \bibinfo{author}{Bruinsma, R.}
\newblock \bibinfo{title}{{Phase-locked spiking of inner ear hair cells and the
  driven noisy Adler equation}}.
\newblock \emph{\bibinfo{journal}{Interface Focus}}
  \textbf{\bibinfo{volume}{4}}, \bibinfo{pages}{20140022--20140022}
  (\bibinfo{year}{2014}).

\bibitem{Spellings:2015be}
\bibinfo{author}{Spellings, M.},  \bibinfo{author}{Engel, M.}, \bibinfo{author}{Klotsa, D.}, \bibinfo{author}{Sabrina, S.}, \bibinfo{author}{ Drews, A.}, \bibinfo{author}{ M and Nguyen, N. H. P.}, \bibinfo{author}{Bishop, K.J.M.}, \bibinfo{author}{Glotzer, S.C.}
\newblock \bibinfo{title}{{Shape control and compartmentalization in active
  colloidal cells}}.
\newblock \emph{\bibinfo{journal}{Proceedings Of The National Academy Of
  Sciences Of The United States Of America}} \textbf{\bibinfo{volume}{112}},
  \bibinfo{pages}{E4642--E4650} (\bibinfo{year}{2015}).

\bibitem{Yeo:2015jz}
\bibinfo{author}{Yeo, K.}, \bibinfo{author}{Lushi, E.} \&
  \bibinfo{author}{Vlahovska, P.~M.}
\newblock \bibinfo{title}{{Collective Dynamics in a Binary Mixture of
  Hydrodynamically Coupled Microrotors}}.
\newblock \emph{\bibinfo{journal}{Physical Review Letters}}
  \textbf{\bibinfo{volume}{114}}, \bibinfo{pages}{188301--5}
  (\bibinfo{year}{2015}).

\bibitem{Uchida:2011ii}
\bibinfo{author}{Uchida, N.} \& \bibinfo{author}{Golestanian, R.}
\newblock \bibinfo{title}{{Generic Conditions for Hydrodynamic
  Synchronization}}.
\newblock \emph{\bibinfo{journal}{Physical Review Letters}}
 \textbf{\bibinfo{volume}{106}},  \bibinfo{pages}{058104--4}
 (\bibinfo{year}{2011}).

\bibitem{Nguyen:2014dl}
\bibinfo{author}{Nguyen, N. H.~P.}, \bibinfo{author}{Klotsa, D.},
  \bibinfo{author}{Engel, M.} \& \bibinfo{author}{Glotzer, S.~C.}
\newblock \bibinfo{title}{{Emergent Collective Phenomena in a Mixture of Hard
  Shapes through Active Rotation}}.
\newblock \emph{\bibinfo{journal}{Physical Review Letters}}
  \textbf{\bibinfo{volume}{112}}, \bibinfo{pages}{075701--5}
  (\bibinfo{year}{2014}).

\bibitem{GuzmanLastra:2016ki}
\bibinfo{author}{Guzman-Lastra, F.}, \bibinfo{author}{Kaiser, A.} \&
  \bibinfo{author}{Loewen, H.}
\newblock \bibinfo{title}{{Fission and fusion scenarios for magnetic
  microswimmer clusters}}.
\newblock \emph{\bibinfo{journal}{Nature Communications}}
  \textbf{\bibinfo{volume}{7}},  \bibinfo{pages}{13519--10}
   (\bibinfo{year}{2016}).

\bibitem{Fily:2012gq}
\bibinfo{author}{Fily, Y.}, \bibinfo{author}{Baskaran, A.} \&
  \bibinfo{author}{Marchetti, M.~C.}
\newblock \bibinfo{title}{{Cooperative self-propulsion of active and passive
  rotors}}.
\newblock \emph{\bibinfo{journal}{Soft Matter}} \textbf{\bibinfo{volume}{8}},
  \bibinfo{pages}{3002--8} (\bibinfo{year}{2012}).

\bibitem{vanZuiden:2016ek}
\bibinfo{author}{van Zuiden, B.~C.}, \bibinfo{author}{Paulose, J.},
  \bibinfo{author}{Irvine, W. T.~M.}, \bibinfo{author}{Bartolo, D.} \&
  \bibinfo{author}{Vitelli, V.}
\newblock \bibinfo{title}{{Spatiotemporal order and emergent edge currents in
  active spinner materials}}.
\newblock \emph{\bibinfo{journal}{Proceedings Of The National Academy Of
  Sciences Of The United States Of America}} \textbf{\bibinfo{volume}{113}},
  \bibinfo{pages}{12919--12924} (\bibinfo{year}{2016}).

\bibitem{Meeussen:2016dl}
\bibinfo{author}{Meeussen, A. S.}, \bibinfo{author}{Paulose, J.} \&
  \bibinfo{author}{Vitelli, V.}
\newblock \bibinfo{title}{{Geared Topological Metamaterials with Tunable Mechanical Stability}}.
\newblock \emph{\bibinfo{journal}{Physical review X}} \textbf{\bibinfo{volume}{6}},
  \bibinfo{pages}{041029--14} (\bibinfo{year}{2016}).
	
\end{thebibliography}


\begin{addendum}
\item[ Supplementary Information] is available in the online version of the paper
 \item We thank L. Bocquet and P. Chaikin for enlightening discussions. This material is based upon work supported by the National Science Foundation under Grant No. DMR-1554724. JP thanks the Sloan Foundation for support. 
  \item[Competing Interests] The authors declare that they have no competing financial interests.
 \item[Correspondence] Correspondence and requests for materials
should be addressed to JP.~(email: palacci@ucsd.edu).
\end{addendum}


\newpage
\begin{figure}
\centering
\includegraphics[scale=1.1]{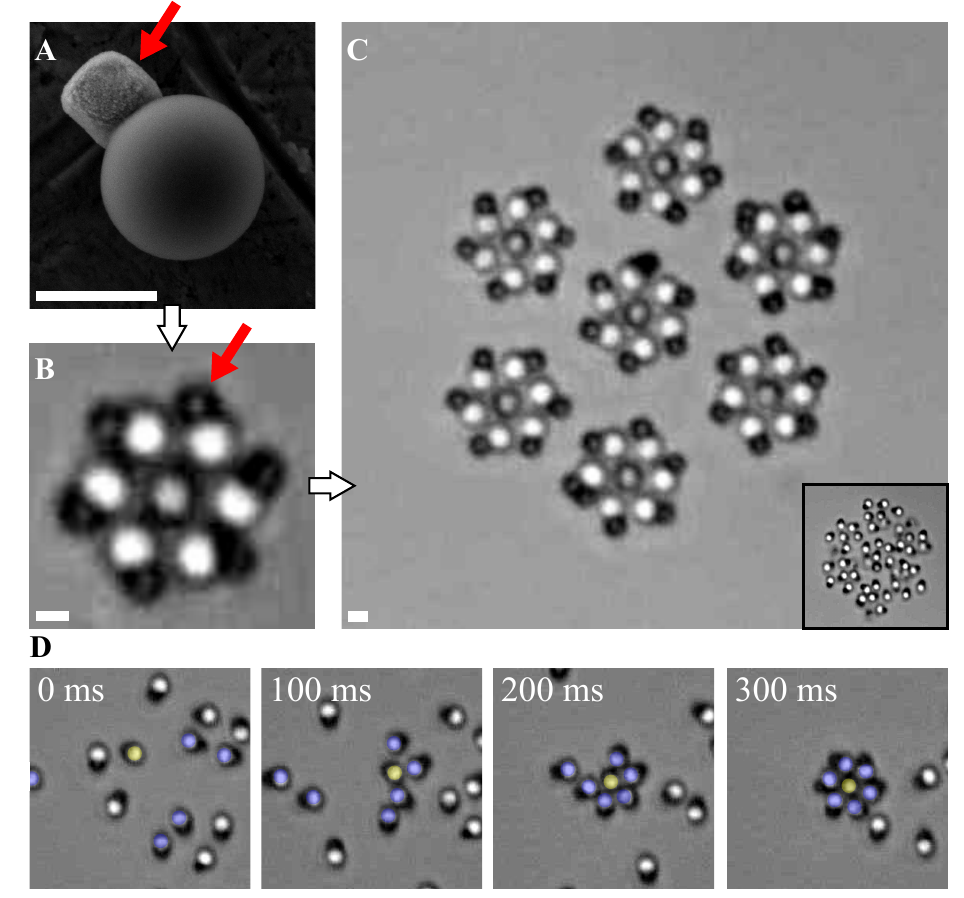}
\caption{\footnotesize\textbf{Hierarchical Self-Assembly of Self-Spinning Rotors}. A) Scanning Electron Microscopy (SEM) of the phototactic swimmer with a fore-aft asymmetry and consisting of the extrusion of the hematite cube  from a chemically inert polymer bead.  The particles self-propel in hydrogen peroxide fuel under light activation, with the bead heading. B) Bright-Field (BF) picture of a self-spinning rotor self-assembled from 7 phototactic swimmers. The core-particle is flipped vertically and surrounded by 6 peripheral swimmers, which orientation defines the  spinning direction. The handedness of the rotor is random, \textcolor{black}{clockwise and counterclockwise rotors are equiprobable}. C) BF imaging of a  dynamical superstructure obtained first by the sequential formation of rotors, then confined thanks to controlled spatiatemporal  light patterns.  C-inset) The self-assembly is purely dissipative, switching off the light, the system returns to equilibrium and the order is destroyed by thermal noise. \textcolor{black}{D) Timelapse of the rapid self-assembly of phototactic swimmers into self-spinning microgear triggered by a focused laser beam. A  particle crosses the high intensity of a laser beam and flips vertically, swimming downwards (yellow particle). It  generates a pumping flow that attracts the peripheral (blue) particles and forms a rotor (see main text). Scale bars are 1 $\mu$m, red arrows indicate the photoactive hematite part.}}
\label{fig1}
\end{figure}

\begin{figure}
\centering
\includegraphics[scale=1]{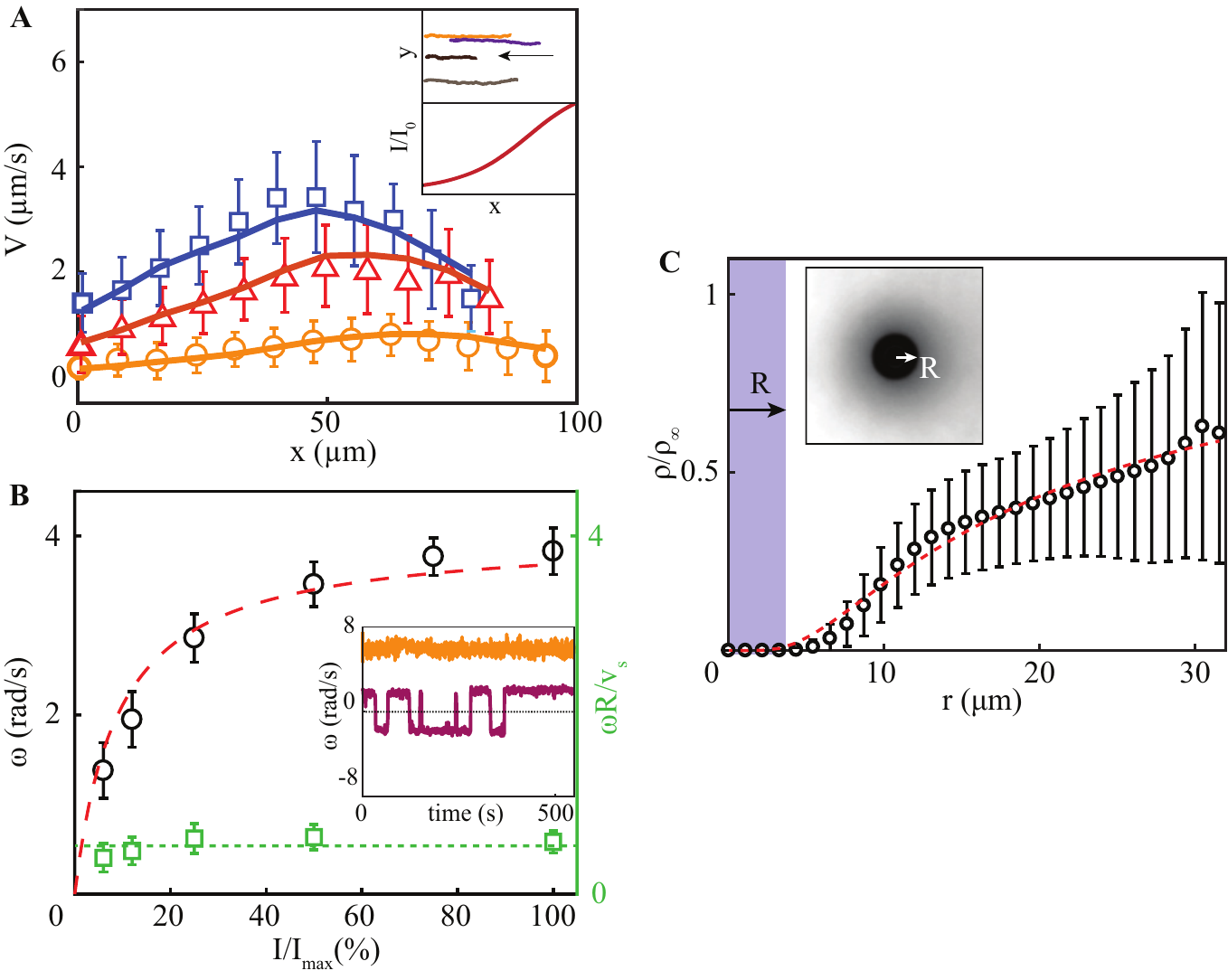}
\caption{\footnotesize\textbf{Light-guided Assembly} A) Migration speed of hematite cubes (empty symbols)  in different light gradients (colors) and comparison with the diffusiophoretic propulsion $V\propto \nabla c\propto \nabla \nu(I)$ (solid lines) obtained from the independent measurement of the intensity profile and a single fit parameter (see Main Text).   A-inset) Example of  particles trajectories (top) and light profile (bottom). Hematite migrates towards the low intensity of light (black arrow).  B) Phototactic swimmers assemble into self-spinning structures rotating at $\omega$, tuned by the intensity of the light (left axis, black circles) reflecting the translational velocity $V_s$ of the phototactic swimmers \textcolor{black}{(red dashed line). The ratio $\omega R/V_s$ is constant  (right axis, green squares)}. B-Inset) Time evolution of the rotation rate $\omega$ for rotors at different speeds. At low speed, $\omega \sim 2$ rad/s,  fluctuations flip the direction of rotation at constant magnitude $|\omega|$ (violet curve). At higher speeds, the rotation is persistent at constant $\omega$ (orange curve). C) Normalized fluorescence showing the repulsion of 200nm fluorescent beads by a rotor. The density $\rho$ of the beads is azimuthally and temporally averaged  from the fluorescence imaging (C-inset) and described by a phoretic repulsion in the gradient of fuel induced by the presence of the rotor: $\rho \propto \text{exp}(-\alpha/rD_\text{c})$ with $\alpha=48\pm 10\mathrm{\mu m}^3$/s (red dashed line). }
\label{fig2}
\end{figure}

\begin{figure}[htbp]
\centering
\includegraphics[scale=1]{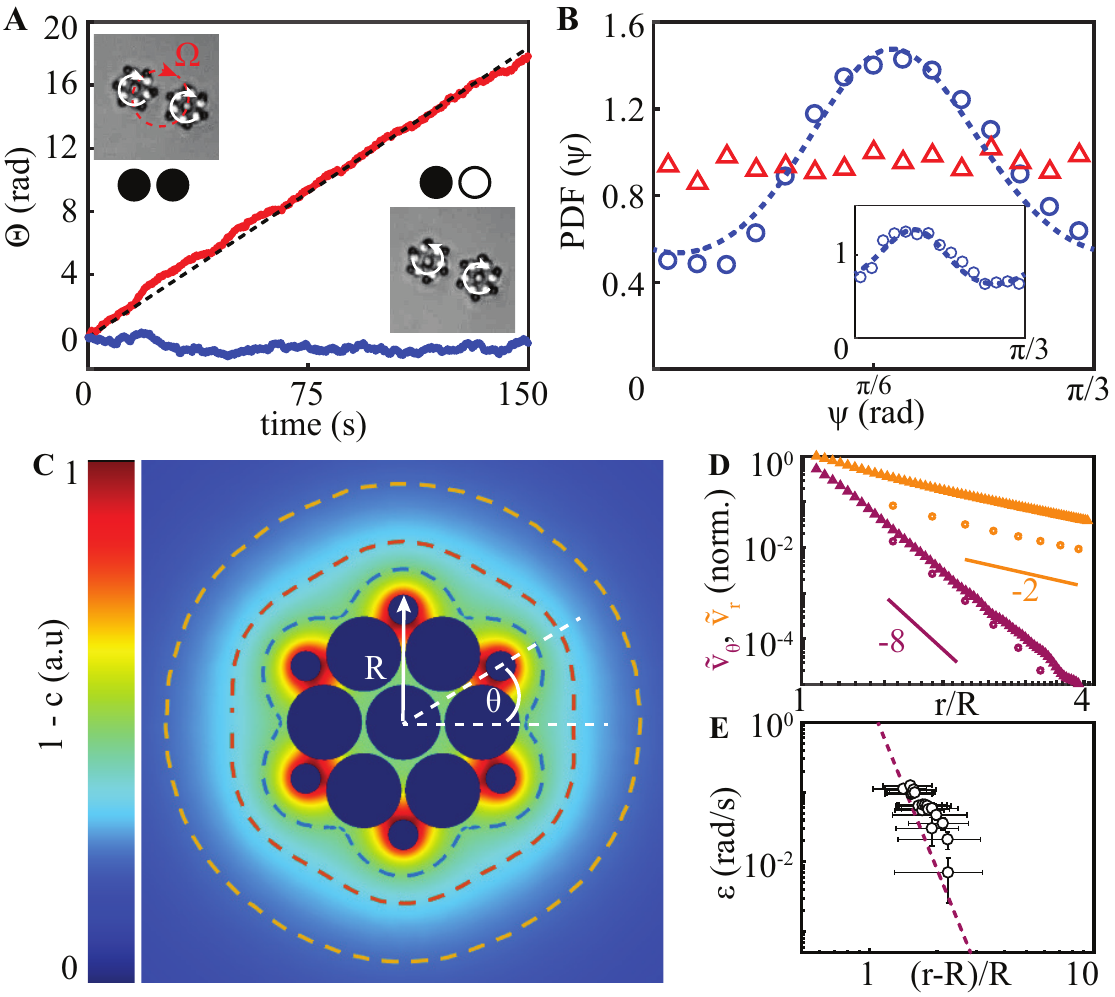}
\caption{\footnotesize\textbf{Rotors Pair-Interactions}. A) Co-rotating pairs revolve at $\Omega\sim 0.1$ rad/s along the common spin (red curve), counter-rotating pairs are static (blue curve).  B) Probability Distribution Function (PDF) of the phase difference $\Psi$ between the rotors. The PDF is flat for co-rotating pairs (red triangles) and shows a peak of synchronization at $\pi /6$  for counter-rotating pairs (blue circles). It is described by a Langevin model of chemically coupled oscillators (dashed line) with the phoretic coupling  $\varepsilon$ as single fit parameter (see Main text). B-inset)  Contra-rotating pairs, with different rotation speeds, exhibit an asymmetric PDF($\Psi$) with a shifted maximum (blue circles), a behavior remarkably captured by the model (dashed line). C)  Simulated concentration of fuel $c$ surrounding a rotor modeled as a structure of 7 impermeable and passive spheres, decorated by 6 chemically-active sites (SI for numerical details).  Dashed lines are isocontours. The concentration field exhibits the 6-fold symmetry of the rotor in the near-field and is isotropic further. It induces the short-range directional interactions between rotors.  D) Radial, $\tilde v_{r}\propto 1/r^2$ (orange symbols), and azimuthal, $\tilde v_{\theta}\propto 1/r^8$ (violet symbols), diffusiophoretic velocities obtained from the simulated concentration field for a point-particle (triangles) and a bead of finite radius $0.3R$ (circles).  E) Phoretic coupling  $\varepsilon$ obtained by fit of the  PDF($\Psi$) (Fig.3B)  for contra-rotating pairs (black symbols), and comparison with the \textcolor{black}{prediction $\varepsilon\sim 2 {\left(\frac{R}{r-R} \right)}^8$ obtained from the radial repulsion, without adjustable parameters (violet dashed line, see Main Text).}}
\label{fig3}
\end{figure}

\begin{figure}[htbp]
\centering
\includegraphics[scale=1.1]{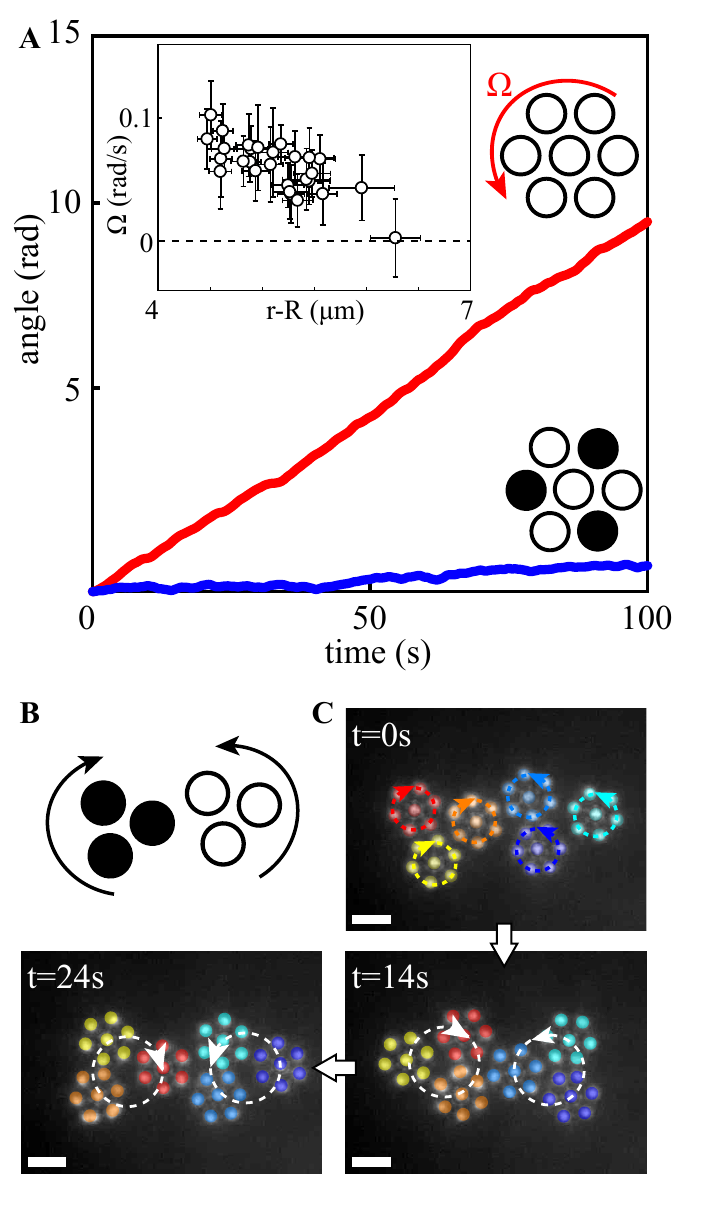}
\caption{\footnotesize\textbf{Dynamical Superstructures}. A) Collections of 7 rotors form a hexagonal  arrangement as a result of radial repulsion and confinement. The dynamics is set by the interplay between spins and spatial coordinates of the components:  co-rotating gears exhibit edge current traveling at $\Omega\sim 0.1$ rad/s along the common direction of spin (red curve), gears with alternating spins are static (blue curve). A-inset) The amplitude of $\Omega$ is controlled by the confinement of the structure as the azimuthal coupling $\varepsilon$ decreases with distance (see Main Text). B) Superstructure made of 2 sets of 3 co-rotating gears, with opposite spins. C) Hierarchical assembly of those two sets into a structure, leads to a synchronous motion of the two sets as visible on the time-lapse pictures. The false colors show the time evolution of the rotors in each set. Black, respectively white,  disks represent clockwise, respectively counter-clockwise,  rotors.}
\label{fig4}
\end{figure}

\end{document}